# Model coupling friction and adhesion for steel-concrete interfaces

**Michel Raous**
Laboratoire de Mécanique et d'Acoustique,
CNRS, Marseille, France
E-mail: raous@lma.cnrs-mrs.fr

**M'hamed Ali Karray**
Ecole Nationale d'Ingénieurs de Tunis,
BP 37 Le Belvédère, Tunis, Tunisie
E-mail: karray_lgc@yahoo.fr

**Abstract:** In this paper the interface behaviour between steel and concrete, during pull out tests, is numerically investigated using an interface model coupling adhesion and friction. This model (RCCM) was developed by Raous, Cangémi, Cocu and Monerie. It is based on the adhesion intensity variable, introduced by Frémond, which is a surface damage variable and its values vary between zero (no adhesion) and one (perfect adhesion). The RCCM model is characterized by a smooth transition from total adhesion to usual Coulomb friction law with unilateral contact. The aim result of this numerical study is the generalization of the RCCM model by taking a variable friction coefficient in order to simulate the behaviour of the steel-concrete interface during a pull-out test. Identification of the parameters and validation of the model are conducted through comparisons of the simulation to experimental results conducted at the INSA of Toulouse.

## 1 INTRODUCTION

The understanding and the modelling of the interfaces is actually one of the challenges of Civil Engineering. It concerns as well the classical technics in masonry (including antique momuments) as the new technics used either in assembling or repearing structures or in developping new reinforced concrete adapted to strong sollicitations. A model coupling adhesion and friction, which may take into account the effect of the viscosity of the interface, is used in the present work in order to model the steel-concrete interface.

A model based on interface damage has been first developed for quasi-static problems in (Raous et al, 1997), (Raous et al, 1999), (Cangémi, 1997) and (Raous, 1999). It describes the smooth transition from a perfect adhesive contact to an usual unilateral contact (Signorini conditions) with Coulomb friction. It is based on a variable characterizing the intensity of adhesion which was first introduced by Frémond (Frémond, 1987) (Frémond, 1988). The RCCM model was used for modelling the fiber-matrix interface of composite matérials and was validated on experiments of micro-indentation of a single fiber conducted at the ONERA (Cangemi, 1997) (Raous et al, 1999).

Using a dynamic formulation, the model was then extended to account for the brittle behaviour occurring when a crack interacts with fiber-matrix interfaces in composite materials (Raous and Monnerie, 2002) (Monerie, 2000). In that case the model was used both for the crack progression and for the fiber-matrix interface behaviour.

The quasi-static formulation was also extended to deal with hyperelasticity in (Bretelle et al, 2001). Mathematical results about the existence of the solutions were given in (Cocou-Roca, 2000) without using any regularization on the contact conditions.

The model was used to study the dynamical behaviour of cohesive masonries in (Jean et al, 1999) and (Acary, 2001).

Through an extension to volumic damage it has been also used in Biomechanics to describe the production of wear particles in bone prosthesis (Baudriller, 2003) or the pull-out of a ligament from a bone (Subit, 2004).

Reinforced concrete is a composite material made up of concrete and steel. These two materials are considered as separate contributors to the overall stiffness and nominal strength of reinforced concrete structures. In fact, both components do interact and the overall structural behaviour is sensitive to the interface behaviour law. Consequently, it is necessary to take into account the behaviour of the steel-concrete interface in any rational analysis of reinforced concrete structures. Experimental study of the interaction between the concrete and a bar subjected to a pull out force (Gambarova, 1988) (Hamouine-Lorrain, 1995) shows that bonds evolve progressively from perfect adhesion to dry friction type.

The RCCM model is extended here for describing the steel-concrete interfaces in reinforced concrete by using a variable friction coefficient which simulates a grinding phenomenon of the interface during the sliding. It is used to simulate the experiments of the pull-out of a steel bar from a concrete specimen conducted in Toulouse (Hamouine and Lorrain, 1995) (Hamouine, 1996) (Karray et al, 2004). First, the identification is conducted on one of the set of pull-out experiments and then the validation of the model is demonstrated by comparing theoretical and experimental results on the other experiments using various geometries and the same materials.

This paper is organised as follows. In Section 2 we first present the RCCM model coupling adhesion to friction and the related thermodynamics basis. The variational formulation and the numerical methods used to solve the quasistatic problem are given in Section 3. The extension of the model and the simulation of the concrete-steel interface are extensively presented in Section 4.

## 2 THE RCCM MODEL

The RCC model (Raous-Cangémi-Cocou) has been first given in (Raous et al, 1997) and then extensively presented in (Raous et al, 1999), (Raous, 1999) and (Cangémi, 1997). It has been extended to the present form (RCCM) by introducing progressively the friction through a given function $f(\beta)$ in (Raous and Monerie, 2002) and (Monerie, 2000). Adhesion is characterized in this model by the internal variable $\beta$, introduced by Frémond (Frémond, 1987) (Frémond, 1988), which denotes the intensity of adhesion. It takes its values between 0 and 1 (0 is no adhesion and 1 is perfect adhesion). The use of a damageable stiffness of the interface, depending on $\beta$, ensures a good continuity between the two contact conditions (initial adhesion and final frictional sliding) during the competition between friction and adhesion.

### 2.1 The model

Contact between two deformable bodies occupying the domains $\Omega^1$ and $\Omega^2$ is considered. **[u]** denotes the gap of displacement between the two solids and **R** the contact force, defined as follows :

$$[\mathbf{u}] = \mathbf{u}^1 - \mathbf{u}^2 \quad (1)$$

$$\mathbf{R} = \mathbf{R}^1 = -\mathbf{R}^2 \quad (2)$$

where $\mathbf{u}^1$, $\mathbf{R}^1$ (respectively $\mathbf{u}^2$, $\mathbf{R}^2$) are related to solid $\Omega^1$ (respectively $\Omega^2$). They are separated into normal and tangential components as follows (where $\mathbf{n}^1$ is the normal vector to the contact boundary of solid 1):

$$[\mathbf{u}] = [u_N] \mathbf{n}^1 + [\mathbf{u_T}] \quad (3)$$

$$\mathbf{R} = R_N \mathbf{n}^1 + \mathbf{R_T} \quad (4)$$

The behaviour of the interface is described by the following relationships, where (5) describes the unilateral contact with adhesion, (6) corresponds to the Coulomb friction with adhesion and (7) gives the evolution of the adhesion intensity $\beta$. Initially, when the adhesion is complete ($\beta = 1$), the interface is elastic as long as the energy threshold $w$ is not reached. After that, damage of the interface occurs and consequently, on the one hand, the adhesion intensity $\beta$ and the apparent stiffnesses $\beta^2 C_N$ and $\beta^2 C_T$ decrease, and on the other hand, friction begins to develop. When the adhesion vanishes totally ($\beta = 0$), we get the classical Signorini problem with Coulomb friction. The contact forces are separated into reversible and irreversible parts. The elastic parts are of course reversible and denoted by $R_N^r$ and $R_T^r$. The model is then written as follows.

- **Unilateral contact (Signorini conditions) with adhesion**

$$-R_N^r + \beta^2 C_N [u_N] \geq 0 \,;\, [u_N] \geq 0 \,;$$
$$(-R_N^r + \beta^2 C_N [u_N])[u_N] = 0 \quad (5)$$

- **Coulomb friction with adhesion**

$$R_T^r = \beta^2 C_T [u_T]$$
$$\| R_T - R_T^r \| \leq \mu\, f(\beta)\, | R_N - \beta^2 C_N [u_N] | \quad \text{and}$$

if $\| R_T - R_T^r \| < \mu\, f(\beta)\, | R_N - \beta^2 C_N [u_N] | \Rightarrow [\dot{u}_T] = 0$

if $\| R_T - R_T^r \| = \mu\, f(\beta)\, | R_N - \beta^2 C_N [u_N] | \Rightarrow$

$\exists\, \lambda \geq 0$, such that $[\dot{u}_T] = -\lambda\, (R_T - R_T^r) \quad (6)$

- **Evolution of the intensity of adhesion**

$$\frac{\partial \beta}{\partial t} = -\frac{1}{b}(w - \beta(C_N [u_N]^2 + C_T [u_T]^2))^- \text{ if } \beta \in [0, 1[$$

$$\frac{\partial \beta}{\partial t} \leq -\frac{1}{b}(w - \beta(C_N [u_N]^2 + C_T [u_T]^2))^- \text{ si } \beta = 1 \quad (7)$$

where $(q)^-$ denotes the negative part of the quantity q.

The constitutive parameters of the model are the following ones.
- $C_N$ and $C_T$ (N/m) are the initial stiffnesses of the interface,
- $w$ is the decohesion energy (as long as that treshold is not reached, adhesion stays to be complete and the behaviour of the interface is elastic with the initial stiffnesses $C_N$ and $C_T$),
- $\mu$ is the friction coefficient,
- $b$ (J.s/$m^2$) is the viscosity associated to the evolution of the adhesion ; we shall see that this effect can be considered as neglectible in the case of concrete.

The term $f(\beta)$ in the friction law with adhesion has been introduced in the Monerie's thesis (Monerie, 2000). It allows introducing the friction progressively when adhesion decreases. This function is such that $f(1)=0$ (no friction when adhesion is complete) and $f(0)=1$ (when adhesion has totally collapsed, friction acts with the friction coefficient $\mu$). We use :

$$f(\beta) = 1 - \beta^m \qquad (8)$$

Raous and Monerie used m=1 for the fiber-matrix interfaces (Raous-Monerie, 2002). In the present work, during the identification of the parameters in the section 4.2, m=2 turns out to be better.

Figure 1 and Figure 2 give the normal and tangential behaviour of the interface during loading and unloading. We set :

$$C_N = C_T = C \ , u° = \sqrt{w/C} \ , R° = \sqrt{wC} \qquad (9)$$

It should be noted that the Signorini conditions are strictly imposed when compression occurs. Regularizations as penalization or compliance are not used; as in the case of the classical Signorini problem with Coulomb friction, we still have multivalued applications both for the normal and the tangential behaviour of the interface.

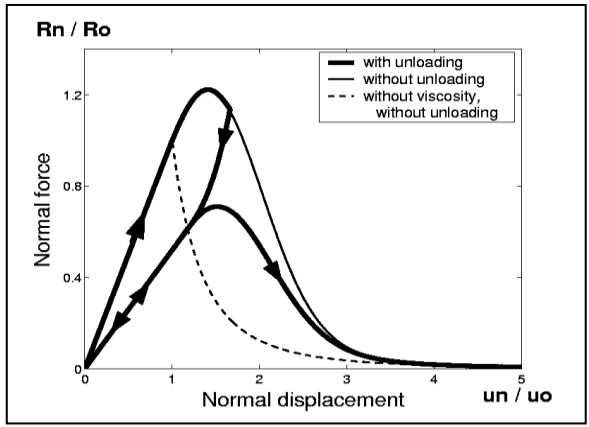

**Figure 1 :** *Normal behaviour law of the interface*

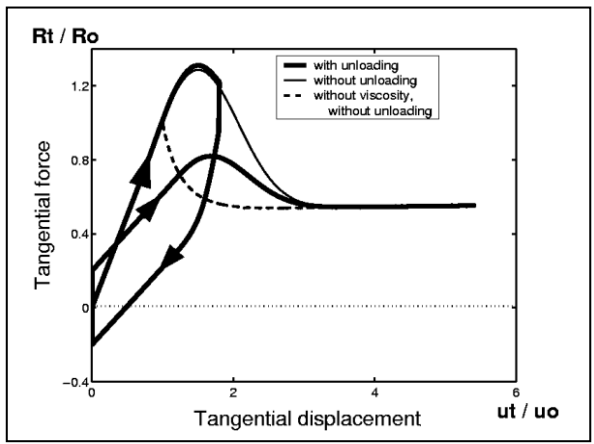

**Figure 2 :** *Tangential behaviour law of the interface*

In section 4, the model will be extended in order to take into account grinding phenomena which may occur during the sliding between steel and concrete : a friction coefficient depending on the sliding displacement will be introduced.

References on other interface models can be found in (Raous, 1999) and (Raous et al, 1999) and a comparison between some of them is presented in (Monerie et al, 1998): they are models developed by Tvergaard-Needleman, Girard-Feyel-Chaboche, Michel-Suquet, Allix-Ladevèze, etc. Using penalization and augmented Lagrangian on a model similar to the RCC one, Talon and Curnier have solved the quasi-static problem using generalized Newton method (Talon and Curnier, 2003).

Specific models for the interfaces in reinforced concrete have been proposed by (Yankelevsky, 1997), (Nago and Scordelis, 1967), (Desir et al, 1999), (Bazant et al, 1995) and (Dominguez Ramirez N., 2005).

### 2.2 Thermodynamics basis of the model

In the framework of continuum thermodynamics, the contact zone is assumed to be a material surface and the local constitutive laws are obtained by choosing the two specific forms of the free energy (10) and of the dissipation potential (11) associated to the surface. Details on the thermodynamic analysis can be found in (Raous et al, 1999), (Cangémi, 1997) and (Monerie, 2000).

The free energy is chosen as follows:

$$\Psi_s([\mathbf{u}], \beta) = \frac{C_N}{2}[u_N]^2 \beta^2 + \frac{C_T}{2}[u_T]^2 \beta^2 - w\beta + I_{R^+}([u_N]) + I_{[0,1]}(\beta) \qquad (10)$$

where the indicator functions $I_{R^+}$ and $I_{[0,1]}$ impose respectively the unilateral conditions and the condition $\beta \in [0, 1]$.

The potential of dissipation is the following:

$$\Phi_s([\dot{u}_T], \dot{\beta}, \chi_N) = \mu\ f(\beta)\ |\ R_N^r - \beta^2\ C_N\ [u_N]\ |\ ||[\dot{u}_T]|| + \frac{b}{2}(\dot{\beta})^2 + I_{R^-}(\dot{\beta}) \qquad (11)$$

where $\chi_N$ denotes the set $([u_N], R_N^r, \beta)$ and the indicator function $I_{R^-}$ imposes the condition $\frac{\partial \beta}{\partial t} \leq 0$ (which means that adhesion can not be regenerated).

It has to be noted that $\Psi_s$ has a part which is convex but not differentiable and another part which is differentiable but not convex with respect to the pair (u, β). $\Phi_s$ is convex but has a part which is not differentiable. Consequently, the state laws and the complementarity laws are then written in the sense of differential inclusions in order to obtain the contact behaviour laws given in section 2.1.

The state laws are written as:

$$\mathbf{R}^r_N \in \partial_{[u_N]} \psi, \; \mathbf{R}^r_T \in \partial_{[u_T]} \psi, \; -G_\beta \in \partial_\beta \psi. \quad (12)$$

where $\partial_z$ denotes the subdifferential with respect to the variable $z$.

The irreversible parts of the contact forces are $\mathbf{R}^{ir}_N = \mathbf{R}_N - \mathbf{R}^r_N$ and $\mathbf{R}^{ir}_T = \mathbf{R}_T - \mathbf{R}^r_T$. The only dissipative processes under consideration are related to the adhesion (both damage and viscosity) and to the friction.

The complementary laws are written as follows:

$$\mathbf{R}^{ir}_N = 0, \; \mathbf{R}^{ir}_T \in \partial_{[\dot{u}_T]} \Phi([\dot{u}_T], \dot{\beta}, \chi), \; G_\beta \in \partial_{\dot{\beta}} \Phi([\dot{u}_T], \dot{\beta}, \chi) \quad (13)$$

## 3 VARIATIONAL FORMULATION AND SOLVERS

### 3.1 The quasistatic problem

As said before, we consider the contact between two elastic bodies lying in the domain $\Omega^1$ and $\Omega^2$. The quasistatic problem can be written as follows with the interface model given in section 2.

The part of the boundary initially in contact is $\Gamma_c$. Boundary conditions are given on $\Gamma_D^\alpha$ ($\alpha = 1, 2$). A volumic force $f^\alpha$ is given in $\Omega^\alpha$ and a surfacic force $\varphi^\alpha$ is given on $\Gamma_F^\alpha$. $K^\alpha$ denotes the elasticity tensor in $\Omega^\alpha$.

**Problem 1 (quasistatic problem):**

Find the displacements $u^\alpha$, the stresses $\sigma^\alpha$ ($\alpha = 1, 2$), the strains $\varepsilon$, and the contact force R such that:

$$\varepsilon = \text{grad}_s u^\alpha \quad \text{in} \; \Omega^\alpha \quad (14)$$

$$\text{div} \, \sigma^\alpha + f^\alpha = 0 \quad \text{in} \; \Omega^\alpha \quad (15)$$

$$\sigma^\alpha n^\alpha = \varphi^\alpha \quad \text{on} \; \Gamma_F^\alpha. \quad (16)$$

$$u^\alpha = 0 \quad \text{on} \; \Gamma_D^\alpha \quad (17)$$

$$\sigma^\alpha = K^\alpha : \varepsilon \quad \text{in} \; \Omega^\alpha \quad (18)$$

and on $\Gamma_c$ (5), (6) and (7) hold (RCCM model for the interface).

### 3.2 Variational formulation

The variational formulation of the quasistatic problem leads to the two variational inequalities (19) and (20) (of which one of them is implicit) coupled with the differential equation (21) given precisely in (7). For the sake of simplicity we only give a formal presentation of the problem. Details on the mathematical formulation and on the functional spaces can be found in (Raous et al, 1999) and (Cocou-Rocca, 2000).

**Problem 2 (variational form):**

Find $(u, \beta)$ such that $u(0) = u_0$ and $\beta(0) = \beta_0$ and for almost all $t \in [0, T]$, $u(t) \in K$ and

$$a(u, v - \dot{u}) + j(\beta, u, v) - j(\beta, u, \dot{u}) + c_T(\beta, u, v - \dot{u}) \geq (F, v - \dot{u}) + (R_N, v_N - \dot{u}_N) \; \forall v \in V \quad (19)$$

$$(R_N, z_N - u_N) + c_N(\beta, u, z - u) \geq 0 \; \forall z \in K \quad (20)$$

$$\dot{\beta} = y(\beta, u) \text{ a.e. on } \Gamma_c, \quad (21)$$

where the initial conditions $u_0 \in K$, $\beta_0 \in H$, $\beta_0 \in [0,1]$ satisfy the following compatibility condition:

$$a(u_0, \omega - u_0) + j(\beta_0, u_0, \omega - u_0) + c_T(\beta_0, u_0, \omega - u_0) \geq (F(0), \omega - u_0) \; \forall \omega \in K.$$

and where:

- $a(u, v) = a^1(u^1, v^1) + a^2(u^2, v^2)$
  $\forall u = (u^1, u^2), v = (v^1, v^2) \in V \quad (22)$

  where $a^\alpha(u^\alpha, v^\alpha) = \int_{\Omega^\alpha} K^\alpha_{ijkl} \varepsilon_{ij}(u^\alpha) \varepsilon_{kl}(v^\alpha) \, d\alpha$, $\alpha = 1, 2$.

- $j(\beta, u, v) = \int_{\Gamma_c} \mu \, f(\beta) \left| R_N - C_N \beta^2 u_N \right| \|v_T\| \, ds \quad (23)$

- $c_N = \int_{\Gamma_c} C_N \beta^2 u_N v_N \, ds$

  and $c_T(\beta, u, v) = \int_{\Gamma_c} C_T \beta^2 u_T v_T \, ds \quad (24)$

- $y(\beta, u) = -\frac{1}{b} [\omega - (C_N u_N^2 + C_T \|u_T\|^2) \beta]^- \quad (25)$

- $(F, v) = \sum_{\alpha=1,2} [\int_{\Omega^\alpha} f^\alpha v^\alpha dx + \int_{\Gamma_F^\alpha} \varphi^\alpha v^\alpha ds] \; \forall v \in V \quad (26)$

- $V^\alpha = \{ v^\alpha \text{ such that } v^\alpha = 0 \text{ a.e. on } \Gamma_C^\alpha \}$, $(\alpha = 1, 2)$, $V = V^1 \times V^2, \quad (27)$

- $K = \{ v = (v^1, v^2) \in V; v_N \geq 0 \text{ a.e. on } \Gamma_c \} \quad (28)$

It has been demonstrated in (Raous, 1999) that the variational formulation of the incremental problem can be written as the following single variational inequality coupled with the differential equation (7). This single variational inequality includes terms evaluated at the previous time step which express the dependance of the solution on the loading path: this is characteristic of the velocity formulation of the friction law.

An incremental formulation is obtained by operating a time discretization of **Problem 2**, taking $n \in N^*$ and setting $\Delta t = T/n$, $t^i = i \Delta t$ and $F^i = F(t^i)$ for $i = 0, \ldots, n$. For the differential equation, we use an implicit scheme. We obtain the following sequence of problems ($P_i^n$), $i = 0, \ldots, n-1$, defined for a given ($u^0, \beta^0$).

**Problem 3 (incremental form):**

Find ($u^{i+1}, \beta^{i+1}$) $\in K \times H$ such that:

$$a(u^{i+1}, \upsilon - u^{i+1}) + j(\beta^{i+1}, u^{i+1}, \upsilon - u^i)$$
$$- j(\beta^{i+1}, u^{i+1}, u^{i+1} - u^i) + c(\beta^{i+1}, u^{i+1}, \upsilon - u^{i+1}) \geq$$
$$(F^{i+1}, \upsilon - u^{i+1}) \quad \forall \upsilon \in K \qquad (29)$$

$$\beta^{i+1} - \beta^i = \Delta t \, y(\beta^{i+1}, u^{i+1}) \quad \text{a.e. on } \Gamma_C, \quad (30)$$

where $c(.) \equiv c_N(.) + c_T(.)$.

The equation (30) is solved using a fixed point method on $\beta^{i+1}$ where the problem (29) has to be solved at each step.

### 3.3 The discretized problem

A fixed point method is introduced on the sliding limit in order to obtain a sequence of Tresca problems. These problems are much more simple to solve because they can be set as minimization problems under constraints (the non penetration condition) of non differentiable functionals. Details can be found in (Raous, 1999). This is written as follows.

**Problem 4 (fixed point on the sliding limit):**

At each time step $t_{i+1}$, find the fixed point of the application S:

$$S(g) = -\mu \, F_N(u_g^{i+1}) \qquad (31)$$

where $u_g^{i+1}$ is the solution of the following **Problem 5.**

**Problem 5 (minimisation problem associated to the Tresca problem):**

Find $u \in k$ such that:

$$J(u) \leq J(\upsilon) \quad \forall \upsilon \in k \qquad (32)$$

with

$$J(\upsilon) = \frac{1}{2} \upsilon^T A \upsilon + G^T |\upsilon - u_h^i| + \frac{1}{2} \upsilon^T C(\beta) \upsilon - F_h^{i+1^T} \upsilon \qquad (33)$$

where:

- $k = \{ \Pi \, K_i \text{ with } K_i = R^+ \text{ if } i \in I_N \text{ and } K_i = R \text{ if not} \}$
- $I_N$ is the set of the number of degrees of freedom concerning the normal components of the contact nodes,
- A is the matrix of dimension N=dim(V): $A_{ij} = a(\omega_i, \omega_j)$,
- C is the diagonal matrix of dimension M (M is the number of contact nodes): $C_M = c(\beta^h, \omega_k, \omega_l)$,
- G is the vector of dimension M: $G_j = \int_{\Gamma_C} g \omega_j \, ds$.

### 3.4 The solvers

In order to solve the minimisation problems various methods have been implemented in the finite element code GYPTIS at the LMA (Latil and Raous, 1991) (Raous, 1999): projected Gauss-Seidel with different accelerations, projected pre-conditionned Conjugate Gradient method (with regulariaztion of the friction in that case). A mathematical programming methods (Lemke) is also used when the problem is formulated as a complementary problem. In the present work, the projected Gauss-Seidel method with an Aitken acceleration has been used. Details can be found in (Raous, 1999) (Raous, 2006).

## 4 SIMULATION OF THE STEEL-CONCRETE BOND BEHAVIOUR DURING A PULL-OUT TEST

### 4.1 The experiments

According to RILEM recommendations (RILEM, 1998), the steel-concrete bond can be characterized by pull out tests. Therefore we test the ability of the RCCM model to describe the behaviour of steel concrete bond during a pull out test. This test consists in the direct wrenching of a steel bar from a concrete specimen as indicated in figure 1. Only a part (with length $l_e$) of this bar is anchored in the concrete specimen, the rest is isolated from the concrete by plastic sleeves as recommended by RILEM and showed in figure 3.

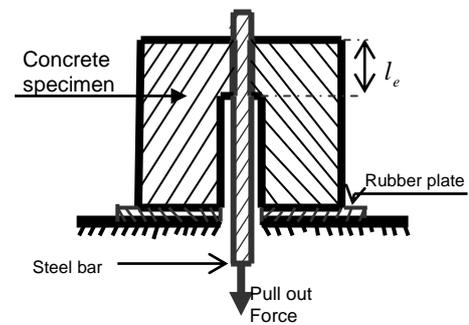

**Figure 3 :** *The experiment*

The results of this test are expressed by a sliding curve representing the variation of the pull out force as a function of the bar sliding. We used the experimental results obtained by Lorrain and Hamouine at the Laboratoire d'Etudes Thermiques et Mécaniques at the INSA of Toulouse where the second author stayed in November 1998 (Hamouine-Lorrain, 1995) (Hamouine, 1996).

These tests are displacement controlled. The concrete specimen considered is a cylindrical concrete block with diameter 160 mm and height 320 mm. The concrete has of age 3 days and its compressive strength is 43MPa. The steel bar is a smooth rod. These tests were performed by all combinations of the following values of the dimensions of the diameter of the bars and of the length of the initial adhesive zone between the bar and the concrete:
- $d$ = 10 mm, 12mm, 14 mm,
- $l_e$ = 50 mm, 100 mm, 150 mm.

All the tests are conducted with a pull out rate of 30 mm/mn. The mechanical characteristics of both materials are given in table 1.

| Material | Young modulus (MPa) | Poisson's ratio |
|---|---|---|
| Concrete | 38000 | 0.2 |
| Steel | 200000 | 0.3 |

Table 1: Mechanical characteristics of materials

This study is conducted in two steps. The first step consists to identify the model parameters and to choose the friction function by using one of the experimental sliding curves: the one corresponding to $d$ = 12 mm and $l_e$ = 100 mm. The second step concerns the validation of the generalized model (RCCM + variable friction) in checking the ability of this model to predict the other experimental results when the parameters identified in the first step are used.

### 4.1 Discretization and loadings

According the symmetries, the computations are conducted under the assumption of cylindrical symmetry. The finite element mesh is given in figure 4. Three nodes triangular elements ($T_3$) are used. The mesh has 1388 nodes of which 21 are contact nodes of the interface.

- **Effect of concrete shrinkage**

Using the cylindrical symmetry hypothesis, the shrinkage deformation, can be written as follows (Zienkiewicz, 1977):

$$\varepsilon_{rr} = \varepsilon_{zz} = \varepsilon_{\theta\theta} = \varepsilon^0$$

where $\varepsilon^0$ is a prescribed deformation. Taking into account the conditions of the concrete specimen, its age and its preservation in a saturated room, the evolution law of concrete shrinkage (BPEL 91, 1994) gives:

$$\varepsilon^0 = -5.6 \cdot 10^{-9}.$$

- **Loading**

Displacement is applied incrementally on the steel bar with a maximum value of 6 mm, divided into 300 increments.

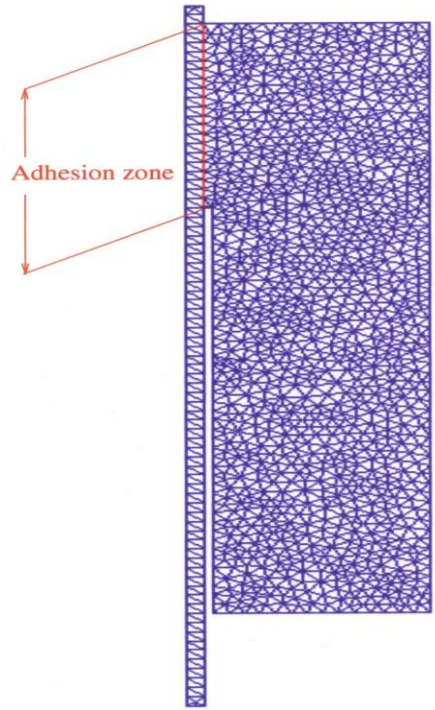

**Figure 4**: *Finite element mesh (1338 nodes)*

### 4.2 Identification of the parameters of the RCCM model

The identification of the parameters is conducted on the simulation of one of the experiments. We choose the case *d=14mm, $l_e$= 10cm* (the diameter of the bar is *14mm*; the length of the adhesive contact between the bar and the concrete is *10cm*). The experimental datas are plotted by the squares on all the following figures.

- **Decohesion energy**

The decohesion energy depends on the concrete age. According to Bazant (Bazant et al, 1995), its value, at early age, can be taken as:

$$w = 100 \text{ J/m}^2$$

- **Initial stiffness**

During pull out tests, the kinematic behaviour of the interface is governed by the tangential component [$u_T$] of the displacement jump. However, the normal component [$u_N$] is not involved. Therefore the value of the normal stiffness $C_N$ has no significant effects. Hence, the same contact stiffness is considered in the tangential and normal direction. It is determined from the initial linear behaviour of the experimental sliding curve (see figure 5, 6 and 8).

$$C = C_N = C_T = 16 \text{ N/mm}^2$$

- **The friction coefficient and the interface viscosity**

Little work has been devoted to the evaluation of friction coefficient µ between steel and concrete during pull out test. In fact, the choice of the values of the parameters µ and b is relatively difficult. For that reason, we conduct first the process of identification of these two parameters on one of the experiments instead to take values from the literature. In the literature, according to the French reinforced concrete code (BAEL 91, 1994), the value of the friction coefficient would be about *0.4*; the precise identification of the coefficient will be conducted in this range of values. Assessment of the viscosity parameter b would require carrying out either tests at different uniform loading rate or tests at constant loading over specified time interval (relaxation or creep). However, these parameters have a significant influence on the shape of the descending branch of the numerical sliding curve (figures 5 and 6). Therefore this couple of parameters can be identified by adjusting the numerical descending branch to that given by the experimental sliding curve. This identification is conducted in two steps as follows.

  *- Constant friction coefficient*

We try in that section to find a set of values (µ, b) convenient to fit the experimental curves. Figure 5 and 6 that, where a constant coefficient of friction is used, show that this is impossible and consequently a variable friction coefficient has to be introduced.

Let us give now some detail on this first tentative of identification using constant parameters. This is conducted on the experimental curve corresponding to *d = 12mm* and *$l_e$ = 100mm*.

By taking into account the effect of confinement resulting from $\varepsilon^0$ and considering a Coulomb friction law, the residual friction coefficient identified from the asymptotic branch of the experimental curve is fist taken as $\mu_r$ = 0.28, which is smaller than the value given in the literature (BAEL91, 1994) . The influence of the viscosity parameter is then studied on Figure 5 by considering the following two values of b:

- b = 0 (no significant viscosity effects)
- b = 130 Js/m$^2$, identified from the maximum pull out force of the experimental sliding curve.

Figure 5 shows that the peak shape of the experimental sliding curve can not be conveniently predicted by varying the viscosity parameter (b = 0 or b = 130 Js/m$^2$). Moreover, the role of the viscosity can be considered as non significant for the concrete on short period of time. Hence, b will be taken equal to zero (b = 0).

In that case, we try now to approximate the experimental curve with different values of the friction coefficient µ. On Figure 6, we give the simulations with:

- µ = 0.45 (in agreement with the values of the literature) which fit the peak but which is not convenient for the asymptotic force,
- µ = 0.28 which is convenient for the asymptotic value of the force but does not fit the peak.

As shown in Figure 6, the descending branch of the experimental sliding curve can not be simulated by RCCM model using constant friction coefficient. A variable friction coefficient has to be introduced. This parameter must decrease from a static value $\mu_s$ to a value corresponding to residual friction after sliding $\mu_r$.

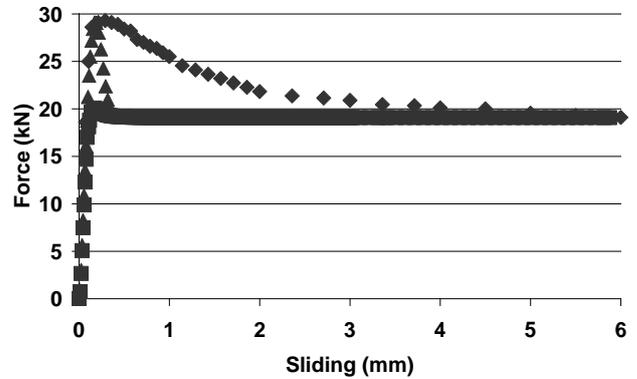

**Figure 5** : *Influence of the viscosity parameter (µ = 0.28):*
- *lozenges : experimental curve d = 12mm, $l_e$ = 10cm;*
- *triangles: numerical results with b=130Js/m$^2$;*
- *squares: numerical results with b=0.*

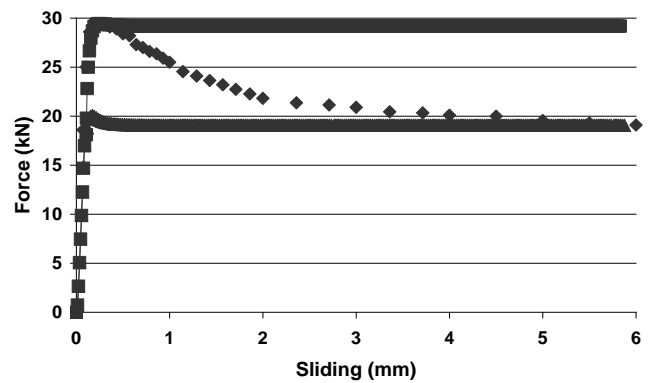

**Figure 6 :** *Influence of friction coefficient µ (b = 0)*
- *lozenges : experimental curve d = 12mm, le = 10cm;*
- *triangles : numerical results for m = 0.28;*
- *squares : numerical results for m = 0.43.*

  *- Variable friction coefficient*

We observed that it was impossible to fit the experimental graph when a constant friction coefficient was used (Karray et al, 2004). That means that, even with various choices for *w, C, b* and *µ*, it was impossible to find a set of values of *w, C, b* and *µ* such that the agreement between the experimental graph and the simulation one was convenient. Regarding the special constitution of the concrete, we now consider that when sliding occurs, dust and small particles are generated inside the interface and that they act as a kind of lubricant: the friction coefficient decreases when sliding occur, a grinding phenomenon appears. To take into account this phenomenon, a friction coefficient depending on the sliding displacement is introduced. We choose the following form for the dependence of the friction coefficient:

$$\mu = \mu_s \quad \text{when } [u_T] \leq 0.3\text{mm} \quad (10)$$

$$\mu = \mu_d + (\mu_s - \mu_d)\, e^{\frac{0.3 - [u_T]}{u_d}} \quad \text{when } [u_T] > 0.3 \text{ mm} \quad (11)$$

where $\mu_d$, $\mu_s$ and $u_d$ are chosen in order that the simulation fit the experimental data as given on Figure 8. We obtained the following values:

$$\mu_s = 0.43; \quad \mu_r = 0.2; \quad u_d = 3\text{mm} \quad (12)$$

The graph is given on Figure 7.

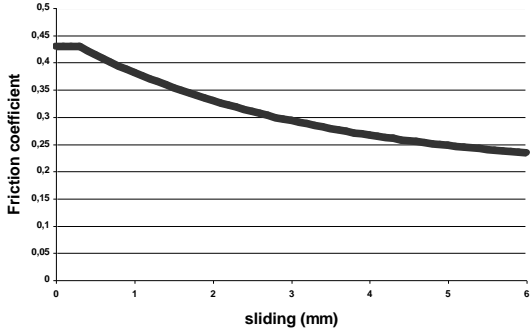

**Figure 7 :** *Grinding law: evolution of the friction coefficient versus the sliding displacement*

The simulation using these values of the constitutive parameters of the model is given on Figure 8 where the full line corresponds to the theoretical simulation and the squares to the experimental results.

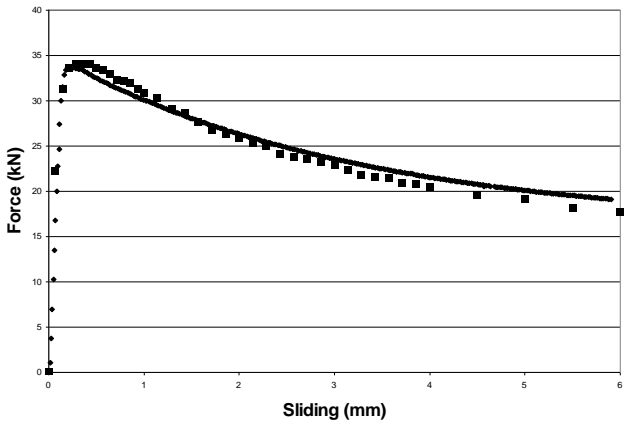

**Figure 8 :** *Identification of the parameters of the RCCM model on the experiment relative to the bar with d=14mm and $l_e$=10cm (full line =RCCM simulation ; squares = experiment)*

The good agreement observed in figure 8 between the experimental and numerical sliding curves only shows that it has been possible to identify a set of parameters and a grinding law for the friction in such a way to simulate conveniently one of the experiments. This was not easy since a variable friction coefficient had to be used but it stays to be of poor interest if the model is not validated on the other experiments when the same values of the parameters are used. That is the aim of the following section.

## 4.3 Validation of the RCCM model on the other experiments

We now simulate all the six different experiments (three rods with different diameters and two different lengths for the initial adhesive contact between the rod and the concrete) using the RCCM model with the values of the parameters evaluated in the section 4.2.

On Figure 9, the results concern three different rods (diameter *d=10* or *12* or *14mm*) with the same initial adhesive contact zone between the rod and the concrete ($l_e$=10cm).

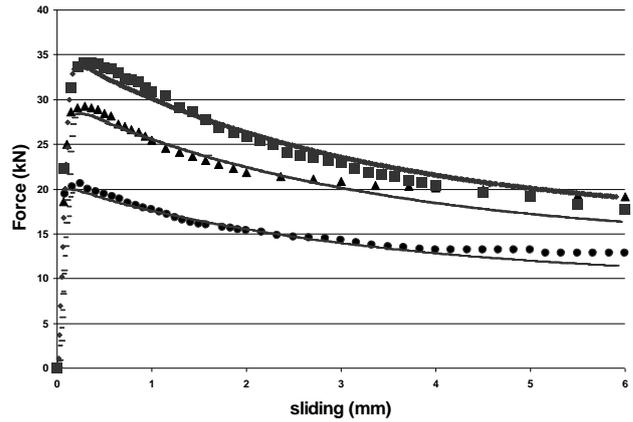

**Figure 9 :** *Validation of the RCCM model: simulation on 3 rods of different diameters (upper results d=14mm; medium results d=12mm, lower results d=10mm ; in all the cases $l_e$=10cm) (full lines =RCCM simulation ; points = experiments)*

We now present the simulation of three experiments conducted on the same geometry of the rod (diameter *d=14mm*) but with three different lengths of the initial adhesive zone ($l_e$= *5* or *10* or *15cm*). Results are given on Figure 10.

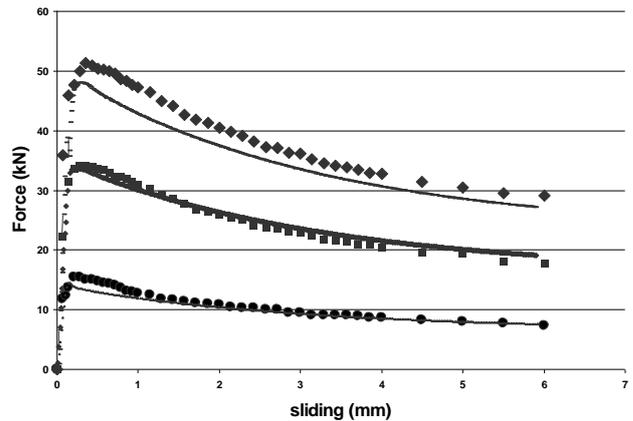

**Figure 10 :** *Validation of the RCCM model : 3 simulation with the same rod (d=14mm) and 3 lengths of the adhesive zone (upper results $l_e$=15cm; medium results $l_e$=10cm, lower results $l_e$=5cm) (full lines =RCCM simulation ; points = experiments)*

## 4 CONCLUSION

A convenient agreement can be observed between the simulations and the experimental results on Figure 9 and on Figure 10. It validates the ability of the RCCM model with a variable friction coefficient (grinding phenomena) to simulate the behaviour of the interface steel-concrete of reinforced concrete for the cases where smooth bars are considered.

An important result of this study is that the numerical prediction of the decreasing branch, of the curve giving the pull out force as a function of the bar sliding, requires to take a variable friction coefficient. Therefore the model is generalized in this work by taking a friction coefficient as a decreasing function of the bar sliding which simulates a grinding phenomenon of the interface. Validation of this generalized model using pull out test data demonstrates its ability to simulate the behaviour of the steel concrete bond with reasonable accuracy.

**Acknowledgement**

We are grateful to Professor Michel Lorrain, INSA, Toulouse, France, and to Doctor A. Hamouine for providing the experimental results given in (Hamouine and Lorrain, 1995) and (Hamouine, 1996).